\documentclass[11pt]{article}
\usepackage{hyperref}
\begin{document}

\title{Immiscible interface vs. miscible interface in Faraday instability}

\begin{author}{\vspace{6pt}W. Batson$^{1,2}$, F. Zoueshtiagh$^{2}$, S. Amiroudine$^{3}$, R. Narayanan$^{1}$
\\ \small{$^{1}$University of Florida, FL 32611 Gainesville, USA,} \\ 
\small{$^{2}$Institut d'Electronique, de Micro\'electronique et de Nanotechnologie UMR CNRS 8520,}\\
\small{Avenue Poincar\'e, 59652 Villeneuve d'Ascq, France,}\\
\small{$^{3}$Laboratoire TREFLE UMR CNRS 8508,}\\
\small{Esplanade des Arts et M\'{e}tiers, 33405 Talence, France}}
\end{author}

\maketitle

%% The abstract (in this file, and that submitted as text to arXiv) should include the exact phrase
%% "fluid dynamics video" or "fluid dynamics videos"

\begin{abstract}
This is a fluid dynamics video submission of miscible and immiscible Faraday instability shot with a high speed camera.
\end{abstract}

% main text

\section{Introduction}

The submitted video depicts the dynamics of a Faraday instability for both a miscible and an immiscible interface. The instability occurs when the system is subject to vibrations of sufficient amplitude at a given frequency. If the amplitude exceeds a critical value, a deflection of the interface appears and grows in time. In the miscible case presented \href{http://arxiv.org/src/1010.2835v1/anc/miscible_vs_immiscible_Faraday_HI.mp4}{here}, water is carefully placed on top of brine and is shaken at a frequency of 3 hz with an amplitude of 40 mm. With surface tension absent, the instability is driven by the density difference of the two phases [1]. What is observed is a subharmonic deflection in response to the forcing frequency, which grows in time until chaotic mixing occurs. In the immiscible case, a 10 cSt silicone oil is placed on top of a fluorinated compound called FC 70, and is shaken at a frequency of 7 hz and an amplitude of 3 mm. While the inclusion of surface tension can help stabilize the deflection to a steady, standing wave pattern, this \href{http://arxiv.org/src/1010.2835v1/anc/miscible_vs_immiscible_Faraday_HI.mp4}{video} shows the shaking can still be violent enough to induce mixing.  Here, the subharmonic deflection grows with each cell period, until the accumulated inertia exceeds what the density difference and surface tension can stabilize. The result is a penetration of the lower fluid into the upper with a mushroom-cloud form, which then breaks further into a 6-fold patterned structure on the resulting upswing of the cell.  Afterwards the interface breaks completely and the system continues to oscillate in a state of chaos.\\

\noindent \small{[1] F. Zoueshtiagh, S. Amiroudine, R. Narayanan, {\it J. Fluid Mech.} {\bf 628}, 43 (2009)}

%\end{enumerate}
%
\end{document}